\documentclass[conference]{IEEEtran}
\usepackage[colorlinks,
linkcolor=red,
anchorcolor=green,
citecolor=blue
]{hyperref} 

\usepackage{amsmath,amssymb,amsthm}
\usepackage{latexsym}
\usepackage{CJK}
\usepackage{cite}
\usepackage{bm}
\usepackage{color, soul}
\usepackage{graphicx}
\usepackage{epstopdf}
\usepackage{epsfig}
\usepackage{subfigure} 
\usepackage{verbatim}  
\usepackage{mathrsfs}	
\usepackage{algorithm} 
\usepackage{algorithmic} 
\usepackage{booktabs}
\usepackage{textcomp}
\usepackage{multirow}
\usepackage{lettrine}
\usepackage[scaled=0.92]{helvet}
\usepackage{mathrsfs}

\usepackage{algorithm}
\usepackage{algorithmic}
\begin{document}

\title{Intelligent networking with Mobile Edge Computing: Vision and Challenges for Dynamic Network Scheduling}

\author{Shuo~Wan, Jiaxun~Lu, Pingyi~Fan,~\IEEEmembership{Senior Member,~IEEE} and Khaled~B.~Letaief{*},~\IEEEmembership{Fellow,~IEEE}\\

\small
Beijing National Research Center for Information Science and Technology(BNRist),\\

Department of Electronic Engineering, Tsinghua University, Beijing, P.R. China\\
E-mail: wan-s17@mails.tsinghua.edu.cn, lujx14@mails.tsinghua.edu.cn, ~fpy@tsinghua.edu.cn\\
{*}Department of Electronic Engineering, Hong Kong University of Science and Technology, Hong Kong\\
Email: eekhaled@ece.ust.hk}

\maketitle

\graphicspath{{Figures/}}

\begin{abstract}
Mobile edge computing (MEC) has been considered as a promising technique for internet of things (IoT). By deploying edge servers at the proximity of devices, it is expected to provide services and process data at a relatively low delay by intelligent networking. However, the vast edge servers may face great challenges in terms of cooperation and resource allocation. Furthermore, intelligent networking requires online implementation in distributed mode. In such kinds of systems, the network scheduling can not follow any previously known rule due to complicated application environment. Then statistical learning rises up as a promising technique for network scheduling, where edges dynamically learn environmental elements with cooperations. It is expected such learning based methods may relieve deficiency of model limitations, which enhance their practical use in dynamic network scheduling. In this paper, we investigate the vision and challenges of the intelligent IoT networking with mobile edge computing. From the systematic viewpoint, some major research opportunities are enumerated with respect to statistical learning.

\end{abstract}

\begin{IEEEkeywords}
Edge computing, Intelligent networking, Statistical learning, Dynamic scheduling
\end{IEEEkeywords}

\IEEEpeerreviewmaketitle

\section{Introduction}\label{Sec:Introduction}

The wide application of internet of things and
cloud services have boosted the need
for edge computing, in which services
occur in some degree at the network edge, rather
than completely in the cloud.
Increasing utilization of network resources is one of the
most key challenges for intelligent mobile network operators.
That is, increasing demand for data communication, cloud services and computation resources.
However, the distributed system can not operate in the same way as centralized system.
In this case, network operators have to increase its intelligence for higher efficiency.
By intelligent networking, edge computing
could address concerns such as processing latency,
limited battery life of mobile devices, bandwidth costs,
security and privacy \cite{yu2018blockchain}.

Data are conventionally collectively managed at the cloud, while most of them are related to mobile services. In such a system,
mobile services rely on collected data and computation resources
hosted in remote cloud. This leads to high network load,
since data have to be up- and downloaded to and from mobile
devices and data centers connected through internet. As pointed out in \cite{nunna2015enabling},
bandwidth demands are expected to continue doubling each year. It is easy to see that the data
traffic in internet of things will overwhelm the current network
capability with time going on due to large variety of network devices and increasing cloud services.
Furthermore, some important emerging network services are typically delay-sensitive.
These issues triggered development of edge computing, where network resources are deployed within the proximity of devices \cite{wan2017smart}. In such a kind of system, intelligent cooperation among edges is a major factor for effective operation.


Computation at mobile devices is limited by battery life
and processor capability. Conventionally, devices will
offload computation tasks to the cloud for assistance.
However, the offloading processing through network may result
in relatively large delay, especially for circumstances with vast mobile devices.
By intelligently offloading computation tasks partly to edges, mobile devices
may obtain a computation assistance services with relatively low latency \cite{wang2017computation,wang2017joint,mao2017stochastic}.

Mobile devices generate huge distributed data commonly with large redundancy.
Therefore, the transmission and processing may be too costly for cloud computing.
By intelligent edge computing, data can be pre-processed to split out redundant part.
The cloud is supposed to receive high quality data with more insights.
Besides, processing data at the network edge
would yield shorter response time,
more efficient processing and less
pressure on the network \cite{kang2018control}.

\begin{figure*}[htbp]
\centering
\includegraphics[width=0.8\textwidth]{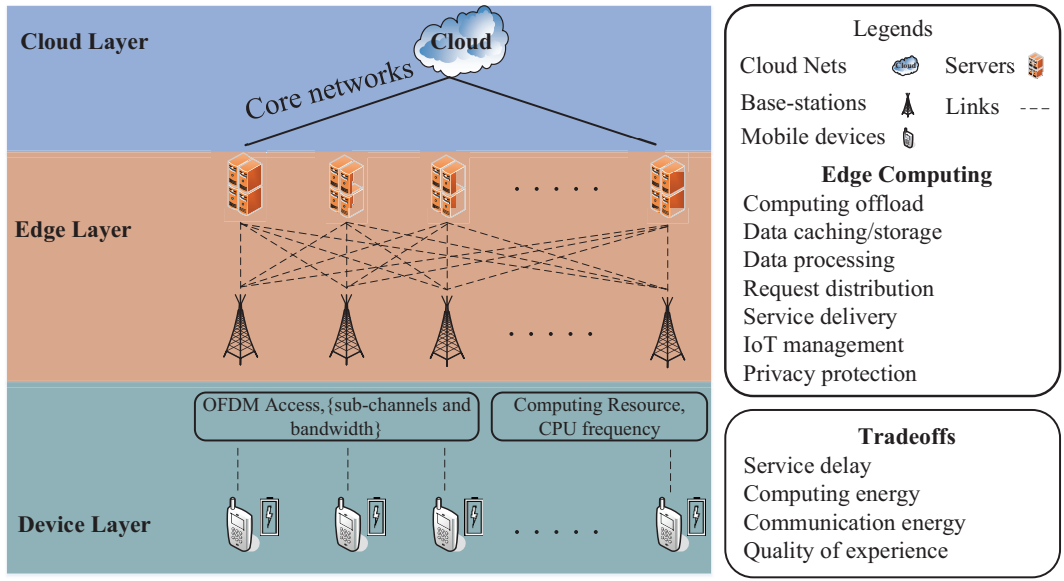}
\caption{Mobile Edge Computing Framework.} \label{Fig:EdgeModel}
\end{figure*}

Edge servers may enhance system efficiency.
However, it also directly affect utilization within the
network core and entail further investments. Rather than managing a single cloud for computation,
the vast edges requires management for cooperation. Both enhancing
existing resources and integrating new technologies comes
with significant operational cost. Consider operation cost among vast network edges,
some works tried to construct system models and optimize the operational cost \cite{park2016joint,you2018exploiting}. However, optimization problem
of cost among multiple objects is typically hard to solve, which may belong to NP-hard problem. Besides, the optimization model may derive some policies in a pre-planning mode, which does not fit the real networking environments with high randomness. Statistical learning may be a
promising technique for adaptive management. Among them, reinforcement learning may be a good representative method for adaptive resource management.
Since it is based on Markov decision process to derive the policy without requiring any specific knowledge of environmental details.
On the simple buffer management, Lyapunov optimization may be a highly efficient tool for dynamic scheduling.

This paper investigates the intelligent networking for mobile edge computing. Based on various critical applications,
the network structure are specified with three layers. Then the mathematical model, typical learning algorithms and research opportunities are introduced.

\section{Intelligent networking Framework}\label{Sec:SysModel}

Fig. \ref{Fig:EdgeModel} shows a general structure of intelligent mobile edge computing (MEC), consisting of three different level layers: device layer, edge layer and cloud layer. \textbf{At the device layer}, mobile devices are limited by battery, storage capacity and computational capability. In this case, it is necessary for mobile devices to upload/download digital resources from clouds, and offload computation tasks to servers, where some intelligent apps are. \textbf{At the edge layer}, MEC servers are deployed inside base stations, and hence they can provide \emph{just in time service} to mobile users. It should be noted that it is not necessary for every base station to be equipped with a MEC server, when taking economic cost into consideration. Instead, edge servers can be installed on macro base stations, which covers a subset of micro cells and provide service to mobile users therein\cite{nunna2015enabling}. In this case, the wireless access points (base stations) and MEC servers are geographically separated and connected via wireless links or fibers. \textbf{At the cloud layer}, MEC servers are connected to remote cloud servers via core networks, which may cause severe delay of services. Therefore, to minimize the computing and data accessing delay, the access to remote cloud servers should be as less as possible. The intelligent networking system for some typical scenarios in IoT are introduced as follows.


\textbf{Data Access Task:} Consider the data accessing task, note that the MEC structure is similar to the CPU hierarchical cache memory, which is widely used to reduce the data accessing delay. In CPU hierarchical cache memory, the accessing delay of L1 to L3 cache increases but the capacity also increases. Hence, the most possible accessed data will be stored at L1-cache, whereas the least possible accessed data should be stored at L3-cache. In considered MEC structure, when considering the data storage case, the mobile device can be considered as the L1-cache, since the access delay and the hinted memory capacity is the smallest. By contrast, the cloud server at the cloud layer can be considered as the L3-cache. Data access with high efficiency is supported by intelligent operation of the layers. Local pattern should be learned by network operator so that useful data can be accessed with lower delay. Different from CPU memory, communications between devices, edges and cloud should also coordinate with the amount of the data transmission. Therefore, it needs an adaptive learner to operate the caching and communication due to the content and amount of locally required data.

\begin{figure}[tbp]
\centering
\includegraphics[width=0.45\textwidth]{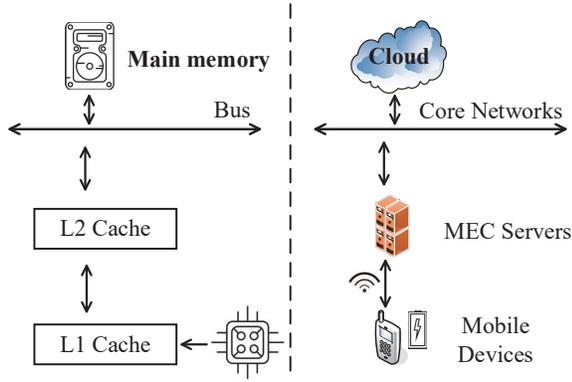}
\caption{Compare the CPU hierarchical cache memory with MEC data acquisition task.} \label{Fig:CPU_MEC_Compare}
\end{figure}

Apart from download, to the upload service, it is essential to reduce the traffic volume at the core network as much as possible. Newly generated data are supposed to be uploaded timely by intelligent scheduling of edge server coverage. Collected data at edges may have large redundancy. Faced with large distributed device data, direct transmission for all raw data may result in huge burden on core network. According to local conditions, edge servers can pro-precess the raw data first, removing some redundancy and shaping the data pattern, or provide some useful results for mobile devices directly. Then the uploaded data to the core cloud may be of high quality with more insights \cite{8908690}. In this process, edge servers and core cloud cooperate on operation of computation and communication. Besides, some overwhelmed edge servers may also offload data to the core cloud or other servers.

It is noted that one important issue in adjusting the data in cache memory is to predict which part of the data is the most possible to be accessed. Similarly, in considered MEC case, one also needs to predict the most possible accessed data.

\textbf{Computing Task:} Consider computing tasks in distributed devices, some works in the literature have shown that using cloudlets to offload computing tasks for wearable cognitive-assistance systems improves response times by between 80 and 200 ms \cite{satyanarayanan2009case} and reduces energy consumption by 30 to 40 percent\cite{shi2016the}. Consider computing servers, the computation and communication resources require dynamic scheduling due to demanded service rate. Besides, the decision of whether to offload computations and which server to offload should be intelligently scheduled from the aspect of devices.

To enhance system performance, the network should smartly predict the intensity of computing tasks generated by mobile users.
The list of works, which need to be done, can be listed as follows:
\begin{enumerate}
\item Predict the mobility pattern of users and track the computing tasks.
\item Model the communication system between mobile users and base stations.
\item Make decision and inference based on previously acquired user activity pattern and communication model.
\end{enumerate}


The intelligent networking mode design also need to consider
device mobility,
high-level dynamics of network topology,
privacy and security protection,
as well as enabling scalability.
Note that centralized networking can not work in IoT network. That is, the intelligent networking must be decentralized due to local learning.

\begin{figure}[tbp]
\centering
\includegraphics[width=0.95\columnwidth]{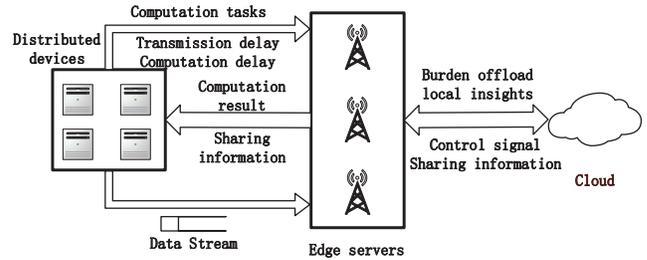}
\caption{Network model for services based on mobile edge computing.} \label{Fig:MEC_MODEL}
\end{figure}

\section{Concept of Networking Model}
In IoT network, the mobile edge computing system poses large sum of servers at the proximity of terminal devices. The smart edge nodes extend the network function beyond simply providing links. Its services involve local data management, computation offloading, distributed data analysis and information sharing, etc. Researchers have set various models for algorithms to support these functions. In this section, some of the major concept of networking models will be introduced briefly.

The networking model of mobile edge computing is summarized as Fig. \ref{Fig:MEC_MODEL}. The network is typically composed of a center, distributed devices and distributed edge servers. The service term mainly consists of edge data and computation offloading. The operation and transmission of the service among network agents is the main objective of the networking model.

From the aspect of networking, a computation task may be characterized by its cost on transmission and computing \cite{wang2017joint}.
The data stream can be quantitatively analyzed by queuing theory \cite{mao2017stochastic}.
Data and computation tasks are generated by underline applications in random mode. The processing capability is dynamically scheduled by intelligent network operator.
The intelligent networking model composed of distributed devices, edge servers and cloud is summarized as follows.

\textbf{Network devices:}
Network devices may offload computation tasks and generated data to edge servers at the proximity. Faced with limited resources and delay requirement, devices require a smart scheduling design due to random environmental elements.

Each term at devices should be decided whether to offload and which edge server to offload. In networking frame, this is typically modeled as a system policy optimized by algorithms. The processor frequency determines the computation delay and energy cost of processors. The communication rate is modeled due to Shannon formula \cite{wang2017joint,mao2017stochastic}. Therein, the system bandwidth, transmission power, noise power and channel power gain are system elements to be optimized with respect to communication delay. The delay within network devices is the sum of transmission delay, server queuing delay and server computation delay. The device should jointly consider power and delay to smartly schedule computation tasks, see \cite{park2016joint,mao2017stochastic,wang2017joint}.


\textbf{Edge servers:}
The edge servers act as local network center with relatively large computation capacity and wide data access. Its entire bandwidth within communications with devices is supposed to be fixed. The bandwidth allocation among devices are dynamically optimized due to service intensity. The objective is to stabilize system and lower down delay within devices, see \cite{mao2017stochastic}.

Computation tasks and uploaded data are temporarily stored in a relay memory for processing. The edge processing management may be modelled by a queue updating in discrete time. The upcoming task is supposed to be random whose average rate can be learned by history information. Its statistical properties are determined by specific user mode. The processing capability is scheduled due to processor frequency and communication resources. With limited power and capability, the processing may be carried out by cloud in part. The offloading rate is characterized by Shannon formula \cite{wang2017joint,mao2017stochastic}. The edge energy is consumed by processing and communication.

The edge servers may cache data in part so that devices could download data with low delay. This process may resemble the caching technique in processor design as shown in Fig. \ref{Fig:CPU_MEC_Compare}. Each content may be analyzed by usage and time. Besides, the edge servers may also learn the user pattern for content usage prediction.

\textbf{Center cloud:}
The edge processed results are transmitted to cloud for further processing. In this case, the cloud may avoid directly collecting raw data from devices, relieving large burden on communications. Besides, the edge servers may also offload part of the raw data faced with heavy burden. It is typically supposed that the cloud has sufficient computation and data resources. Therefore, it is able to answer received requests without limitation on resources.

The services form queues from separate edge servers. These queues may vary largely due to varied local conditions. From the aspect of cloud, the resources should be dynamically scheduled among edge servers. The networking design objective is to stabilize the overall average queuing length in a relatively low level, where each edge server may be allocated a priority. Note that communications between edges and cloud may be wired or wireless. In wireless conditions, the cloud should dynamically allocate bandwidth according to edge service requirements, see \cite{8908690}.

\section{Intelligent Networking: Statistical learning}
In various application scenarios, researchers set corresponding models for network scheduling and solve them by optimization. The conventional strategy tends to derive the policy with a typical scenario. However, the practical network may encounter various complex environments, being much different from the discussed scenarios. The statistical properties within the whole network are typically unknown, which is challengeable for optimization.


How to make the network scheduling smart or intelligent? Statistical learning may be a good way to implement it. As we know, statistical learning is a set of methods for adaptive policy maker. In training process, training data from a specific scenario are learned statistically by a learner. By iterative training on various data, the learner gradually adapts to the application. In scenarios where the specific system is hard to model mathematically, the learner will find a way to fit the system details by iterative learning. Some historical successful applications are observed in picture processing, voice recognition, etc. The key characteristics in the applications is that having not any accurate mathematic model to characterize the varying of object pixels and sound.

Similar phenomenon exists in intelligent networking. IoT network faces distributed agents with high randomness, which requires a dynamic mode. Since the network conditions varies with time, the statistical model should have the capability to learn from history and make scheduling with its observed current conditions. In this section, as an alternative, we will introduce the commonly applied reinforcement learning and Lyapunov optimization.

\subsection{Reinforcement learning}
Reinforcement learning is a major branch of machine learning for policy determination of agents. The agents are supposed to select actions within the complex environment so that the accumulated reward is maximized. Different from greedy algorithm, it concerns estimation of the accumulated future reward rather than simply current reward. Reinforcement learning is typically applied in complex environments without model for explicit actions. By iterative interaction with around environment, the intrinsic environmental elements are learned for action policy. The learning model is typically depicted by Markov decision process (MDP) for dynamic programming.

\begin{figure}[tbp]
\centering
\includegraphics[width=0.9\columnwidth]{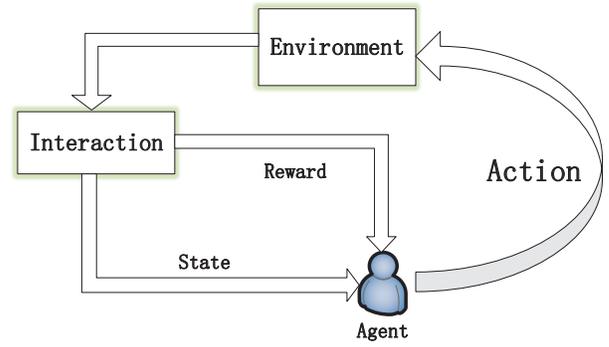}
\caption{Markov decision process (MDP) model.} \label{Fig:MDP_Structure}
\end{figure}


The Markov decision process commonly models dynamic decision making as shown in Fig. \ref{Fig:MDP_Structure}. An agent is supposed to accomplish specific task in a complex environment. Given current environmental state, it selects an action in candidate set by current policy. The environment feeds back the reward and its intrinsic state is updated. Based on the received feedback, the agent trains its policy to fit the environment.

In rather complex network scheduling circumstances where the pattern of state transferring is commonly unknown, Q-learning is typically applied. The Q-function represents the estimated rewards of taking an action in a given state. By feedback of rewards and states, the derived samples of results are utilized to train the estimator of Q-function. The major issue is the approximation of Q-function. The rather simple methods involve kernel function and Q-table. To cope with more complex states in dynamic network, a neural network is trained as Q-function to estimate the reward for separate actions, say deep Q-learning. The system reward can be derived by feedback of actions. By iterative interaction and training, the neural network can be updated to fit the underlying network conditions. The outcome choice of action can gradually enhance the system reward.

In network system, each server is aware of its resources and the conditions of around environment. Each device may know its around servers and its own service requirement. The Q-network can be trained to intelligently choose networking actions based on the locally observed information. By dynamic interactions and training, the edges may approach an optimal policy to deal with the observations. In this way, the distributed edges can dynamically operate for cooperations. Besides, the data for training can be derived by feedback of rewards. Therefore, the deep Q-learning overcomes the controversial requirement for labeled data and better fits the dynamic network.


\subsection{Lyapunov optimization}
System stableness is one of the major problems in intelligent networking. Distributed servers may have varied burden with high randomness, which largely risks the system stableness. Therefore, the resources should be balanced among edges so that the overall system remain stable. In low rate cases, the edges are supposed to reduce power consumption. In high rate cases, edges should increase service capability with more resources of computation and communication.

Lyapunov optimization is commonly applied in control theory to maintain system stableness. The system state is depicted by Lyapunov function. It is typically desirable to achieve a relatively small Lyapunov function. The Lyapunov drift is defined to represent the stableness of Lyapunov function, which is typically the stableness of the whole system.

In dynamic scheduling, the system should take actions so that Lyapunov drift approaches zero in long time range. Furthermore, the cost penalty can be added into Lyapunov drift to jointly optimize system stableness and cost. In network scheduling, Lyapunov optimization can control the system without knowing statistical properties of service requirements. The actions can be determined based on current observations of system state, which achieves an online manner. In this sense, it is a very powerful tool for smart network design.

Suppose services at separate edges are stored in a queuing network. The queuing length in edges may be viewed as system state. The queuing length are affected by randomly upcoming services and varied processing capability. Since we do not know statistical properties of networking environment, Lyapunov optimization is applied to maintain the system stableness by dynamic scheduling, see \cite{mao2017stochastic}.

%

\begin{figure}[tbp]
\centering
\subfigure[Edge power consumption in low service intensity case. ] { \label{power_r}
\includegraphics[width=0.8\columnwidth]{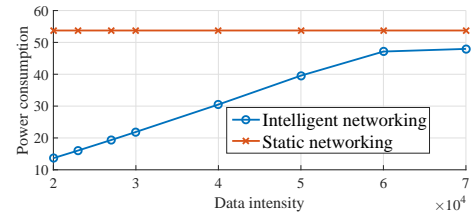}}\\
\subfigure[Data bits staying at distributed edges waiting for processing.]{ \label{rate_r}
\includegraphics[width=0.8\columnwidth]{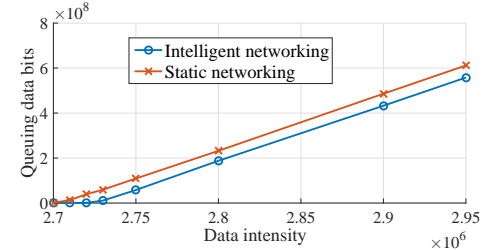}}
\caption{Performance of intelligent networking and static networking.} \label{system_test}
\end{figure}

\subsection{Performance validation of intelligent networking}
As discussed in previous sub-sections, statistical learning is an efficient method for intelligent networking. In this sub-section, the performance of statistical learning method is validated by simulations on Python.

Edge servers are supposed to be distributed in $7$ regions with different service intensity. Data service amount is supposed to satisfy Poisson distribution. The average rate is randomly generated by uniform distribution around a previously set rate $\lambda$. The edge processing frequency is upper bounded by $f_{max}=1.5GHz$. Part of the services are transmitted to cloud for offloading. The transmission rate satisfies Shannon formula. The Lyapunov optimization is applied here to schedule offloading policy, edge processor frequency and transmission power, so that the system can cope with various service intensity with high efficiency. The intelligent networking is compared with conventional static networking, where processing resources are fixed for edges.

Simulations show that intelligent networking can largely reduce power consumption in low service intensity case. As shown in Fig. \ref{power_r}, intelligent networking may lower down its processing capability to save energy in low service intensity cases. As more services are arriving at edges, edges could dynamically increase the processing capability, which is reflected by the increasing energy consumption.

In high rate cases, intelligent networking may also obtain a lower delay. Fig. \ref{rate_r} shows the average queuing data bits in distributed edges. By intelligent networking, resources can be smartly distributed among edges with various service intensity. However, static network may perform worse to cope with various intensity among edges. As Fig. \ref{rate_r} shows, intelligent networking achieves lower level of queuing services, which means lower serving delay.


\section{Major research challenges}
In edge computing network, edge servers act as platforms for cloud services and data processing. The system makes prerequisites for various applications with high-level requirement for delay and flexibility. The typical applications may involve connected cars, smart city and smart grids \cite{ning2019vehicular}.

%

In the above sections, the model and typical learning algorithms are introduced for intelligent networking. In this section, insights of some of major challenges and opportunities in intelligent networking will be discussed as follows.


\subsection{Mobile Coverage Scheduling}
In applications such as connected cars, devices are supposed to move with high mobility, which may not be supported by static base stations. For applications such as smart city, the network may face devices with extremely wide distribution. In this case, researchers start to investigate servers with mobility, which is represented by the Unmanned Aerial Vehicle Base stations (UAV-BSs).

The network servers with mobility enables a flexible coverage of devices. They may have higher efficiency and lower delay in serving distributed devices with high mobility. However, the mobility pattern also increases the complexity of coverage scheduling. A representative issue is the coverage of UAV-BSs.

\begin{figure}[tbp]
\centering
\includegraphics[width=0.9\columnwidth]{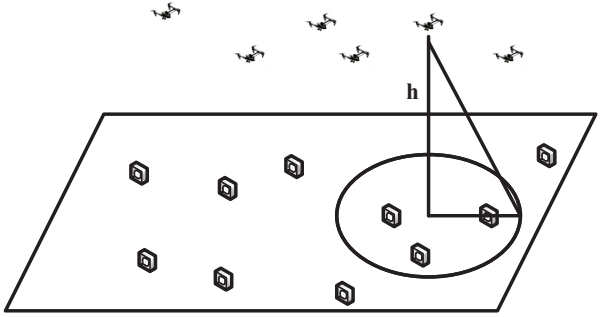}
\caption{Mobile coverage model of UAV base stations for distributed devices.} \label{Fig:UAV_coverage}
\end{figure}

As shown in Fig. \ref{Fig:UAV_coverage}, a disk area on the ground is covered by a UAV base station. In \cite{matolak2015unmanned}, the authors proposed a communication channel for UAV-ground communications. Consider the power and communications, the covering radius and hovering height is optimized by environmental blocking coefficient. For intelligent cooperation among multiple UAVs, the positions are supposed to be dynamically scheduled to ensure an efficient coverage.

The major problems involve mobile path planning, hovering height scheduling and dynamic resource scheduling. By path planning, the mobile servers can adjust their coverage spots. The objective is to achieve a sufficient coverage and usage of the processing resources. The height may influence the service quality within the coverage range, which should be scheduled by environmental elements and conditions of nearby servers. As for service scheduling of mobile servers, it should be dynamically modified according to service demand of the current covered area. Besides, distributed servers may also interact with each other for service offloading.

The coverage scheduling with mobility for multiple servers is a complex issue. The cooperations and interactions therein is very challengeable. We may not have a solvable mathematical model of such system, since the mobility and coupling influence from varied servers make things very complex. In this case, statistical learning method can be applied to learn the coverage policy. Among them, a representative solution is reinforcement learning. Due to networking objectives, we can define the corresponding rewards, actions, state observations and learning model. By adaptive learning, the coverage scheduling policy can be carried out in dynamic mode.

\subsection{Cooperation within Multi-agent system}
The dynamic scheduling for a central server with distributed users have been widely researched. However, in IoT network based on mobile edge computing, the basic system should be composed of devices, edges and cloud. Therefore, the intelligent networking should consider various edge servers and distributed devices. The cooperation and interaction among various edges in serving distributed devices rise up as a crucial issue.

The intelligent scheduling in multi-agent system is a newly raised subject. It aims at resolving dynamic cooperation of multiple agents given a specific task. Intelligent networking for various edge servers is a challengeable issue. The correlated channel model, randomness of device state and interference from other edges are all critical issues. Reinforcement learning can be applied here to learn an adaptive policy maker. It is supposed that each edge could maintain its local observation. The trained policy is supposed to process the observations to obtain a temporary optimal policy. For specific environmental conditions, the corresponding observations should be carefully designed, which involves the essential available information of around environment. The reinforcement learning for multi-agent system is also a popular researching field. Some works proposed to pose training process in central cloud. Others added modifications on sample weights to avoid overtraining on part of the candidate actions. In the field of intelligent networking, the detailed design of reinforcement learning within multiple edges is a promising research field.

\subsection{Resource Scheduling for random service requirements}
High randomness within distributed devices is a major issue for network scheduling. In IoT network, most of the distributed service requirements are generated by machines, which is hard to obtain the corresponding probability distributions. This results in a time-varying burden in separate edges. The edge servers are limited by onboard processing capability. Faced with high service rate, they may be overwhelmed and collapsed. In this case, the communication and computation resources within the network should be scheduled among edge servers, so that edges with large burden may share more resources. Edges may share resources with nearby companions by inter-edge communications.

The network resources mainly involve communications and computations. By communications, edges can offload part of the data or computations to other edges or the cloud. Furthermore, the edges should take power into considerations. In low rate cases, they may lower down the processor frequency to reduce energy consumption.

All the above process require a dynamic policy for random service requirements from distributed devices. Lyapunov optimization is a commonly used technique to remain stableness faced with unknown random environmental events. By jointly considering cost penalty, the resource scheduling policy can be optimized in an online manner. The major problem is to properly define the Lyapunov function and penalty cost. Then the problem can be solved for separate $t$ so that the policy can be derived by current conditions of system stableness.

\section{Conclusion}\label{Sec:Conclusion}
The intelligent networking for IoT network was investigated based on mobile edge computing. The major issues within IoT network may involve data and computation services for vast distributed devices of various types. Consider the complex environmental elements with high randomness, the edges face great challenges in decentralized networking, which should be in online manner. In this paper, we introduced the basic system model and discussed some major statistical learning methods for intelligent networking. Some major problem in intelligent networking were summarized as mobile coverage, multi-agent cooperation and resource scheduling. By detailed introduction of the major challenges, the research opportunities within intelligent networking were proposed. The core idea is to design a smart network for IoT applications in edge computing based on statical learning.

\section*{Acknowledgment}

This work was partly supported by the China Major State Basic Research Development Program (973 Program) No. 2012CB316100(2) and National Natural Science Foundation of China (NSFC) No. 61771283.




\end{document}